\documentclass{article}
\usepackage{amsmath}
\PassOptionsToPackage{numbers}{natbib}
\usepackage{natbib}
\usepackage{graphicx}
\usepackage{subcaption}
\usepackage{titlesec}
\usepackage[font=small,skip=1ex]{caption}  
\titlespacing*{\subsection}{0pt}{0.5em}{0.5em}
\usepackage{tikz}
\usetikzlibrary{shadows}
\usepackage{marvosym} 

\usepackage[preprint]{neurips_2025}

\usepackage[utf8]{inputenc} 
\usepackage[T1]{fontenc}    
\usepackage{hyperref}       
\usepackage{url}            
\usepackage{booktabs}       
\usepackage{graphicx}       
\usepackage{amsfonts}       
\usepackage{nicefrac}       
\usepackage{microtype}      
\usepackage{xcolor}         
\usepackage{algorithm}
\usepackage{amsmath} 
\usepackage{etoolbox}

\usepackage{algpseudocode}  

\title{S-Crescendo: A Nested Transformer Weaving Framework for Scalable Nonlinear System in S-Domain Representation}

\author{%
Junlang Huang\textsuperscript{1,*},
Hao Chen\textsuperscript{1,*},
Li Luo\textsuperscript{1,\dag},
Yong Cai\textsuperscript{1,\dag},
Lexin Zhang\textsuperscript{1},
Tianhao Ma\textsuperscript{1},
Yitian Zhang\textsuperscript{1},\\
Zhong Guan\textsuperscript{1,\Letter}\\ \\
}

\begin{document}

\maketitle

\begin{abstract}
 Simulation of high-order nonlinear system requires extensive computational resources, especially in modern VLSI backend design where bifurcation-induced instability and chaos-like transient behaviors pose challenges. We present S-Crescendo - a nested transformer weaving framework that synergizes S-domain with neural operators for scalable time-domain prediction in high-order nonlinear networks, alleviating the computational bottlenecks of conventional solvers via Newton-Raphson method. By leveraging the partial-fraction decomposition of an n-th order transfer function into first-order modal terms with repeated poles and residues, our method bypasses the conventional Jacobian matrix-based iterations and efficiently reduces computational complexity from cubic $O(n^3)$ to linear $O(n)$.The proposed architecture seamlessly integrates an S-domain encoder with an attention-based correction operator to simultaneously isolate dominant response and adaptively capture higher-order non-linearities. Validated on order-1 to order-10 networks, our method achieves up to 0.99 test-set \(R^2\) accuracy against HSPICE golden waveforms and accelerates simulation by up to 18\(\times\), providing a scalable, physics-aware framework for high-dimensional nonlinear modeling.
\end{abstract}

\section{Introduction}
%
%
%
%
In recent years, deep learning technologies have made remarkable progress, with Transformer-based architectures demonstrating exceptional performance and significant advantages in fields such as natural language processing (NLP), computer vision, and time-series data modeling.\cite{vaswani2017attention} Transformer models, by effectively capturing complex relationships and long-range dependencies, offer a novel perspective for data-driven modeling. This technological advancement has inspired researchers to explore its potential applications in traditional engineering domains, especially in complex physical modeling and signal prediction\cite{devlin2019bert}\cite{schneider2023survey}.

\begin{figure}[h]
    \centering
    \includegraphics[width=1\linewidth]{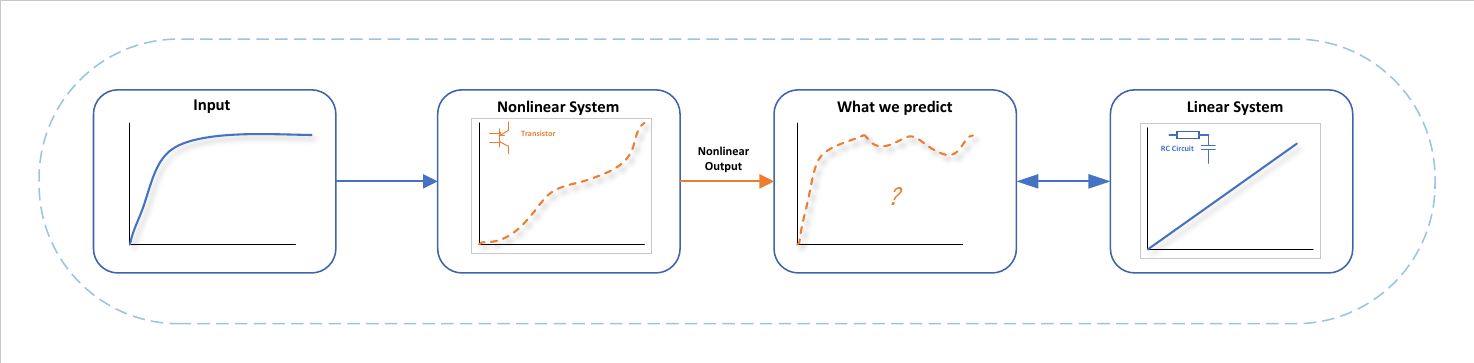}
    \caption{Given a known input signal, the task is to predict the nonlinear system’s output before it feeds into the linear system. This intermediate signal, marked in red, is unknown and serves as the prediction target of our model.}
    \label{fig:enter-label}
\end{figure}

Nonlinear system identification remains a core challenge across many domains, particularly under high-order dynamics, non-stationarity, and limited observability. Classical methods from control theory—such as Volterra series, Hammerstein–Wiener models, and grey-box approaches—typically decompose system behavior into a linear core and a nonlinear correction \cite{antoulas2005approximation, cen2011grey, schetzen1981nonlinear, wills2013identification}. However, their scalability and generalization degrade in high-dimensional parameter spaces \cite{schoukens2019nonlinear}. These limitations are especially pronounced in modern integrated circuit design, where nonlinearities emerge not only from active devices but also from parasitic effects in passive interconnects. A canonical example is the ``nonlinear driver + linear RC load'' configuration, illustrated in Figure~\ref{fig:enter-label}. As technology scales, interconnect parasitics exhibit strong dynamic nonlinearities due to proximity effects, process variation, and material inhomogeneity \cite{sharma2011vlsi}, complicating accurate modeling. To address this, we propose a neural operator framework that integrates Laplace-domain physical priors with data-driven adaptability. Demonstrated on RC current response tasks, this method extends naturally to systems governed by partial differential equations with nonlinear boundary conditions, offering a scalable and physically consistent approach to complex system identification.

In response, a growing body of research has applied deep learning to nonlinear system identification. Approaches based on recurrent networks\cite{schussler2019deep}, attention mechanisms\cite{raissi2019physics}, and neural operators\cite{chen2018neural} (e.g., DeepONet\cite{lu2021learning}, FNO\cite{li2020fourier}, and PINNs) have shown promise, particularly in low-dimensional settings or for PDE systems with known boundary conditions. Yet, these methods often require large datasets, lack robustness to structural variation, and suffer from poor physical interpretability\cite{lundby2023deep}\cite{pillonetto2025deep}. More critically, few methods explicitly leverage the modal structure or transfer function representation intrinsic to many engineered systems. As a result, they remain constrained by either computational inefficiency (due to iterative solvers like Newton--Raphson) or limited generalization to systems with unseen topologies or higher-order dynamics\cite{revay2023recurrent}\cite{lundby2023deep}\cite{andersson2019deep}\cite{niu2022deep}.While deep learning has been explored for general system identification, its application to signal-line RC response modeling—a canonical high-order nonlinear problem—remains largely unexplored. Most existing methods still fall under three classical paradigms: Current Source Models (CSMs) \cite{croix2003blade,synopsysCCS}, Voltage Response Models (VRMs) \cite{qian1994modeling}, and Direct Waveform Prediction \cite{jain2011accurate}. Each faces practical limitations: CSMs lack waveform fidelity due to fixed capacitance abstraction \cite{amin2006multi}, VRMs suffer from high cost and solver-induced errors \cite{abbaspour2003calculating,jiang2010non}, and direct fitting fails under sharp transitions due to overshoot and undershoot distortion.

Against this backdrop, Transformer-based methods offer a powerful new tool for nonlinear system identification in the S-domain. Transformer architectures excel at capturing complex, long-range dependencies and higher-order interactions, making them ideally suited to address the limitations of traditional RC-network models in time-domain response prediction \cite{vaswani2017attention}. 

Even in nominally high-order interconnects, only a handful of “dominant poles” govern signal behavior. Standard model order reduction (MOR) selects these modes, typically fewer than ten by energy or gain thresholds, and collapses a 5000th order RC network into a low-order surrogate with minimal frequency response error \cite{odabasioglu2003prima,cheng2012model}. Going beyond linear reduction, we encode each reduced first-order term into a learned latent embedding and apply an attention-driven correction operator to capture nonlinear driver-load effects. On 4th to 10th order surrogates, our method attains $R^2$ up to 0.99 against HSPICE waveforms, evidencing superior accuracy. By leveraging partial-fraction decomposition, we decouple topology from prediction-forecasting each modal response independently in the S-domain before summation, thus ensuring universality and structural agnosticism.We introduce a physics awared S-domain neural operator that seamlessly integrates with MOR pipelines, delivering a scalable, accurate, and efficient solution for nonlinear RC simulation, electromagnetic analysis, and signal-integrity evaluation without fixed driver or load models.

\begin{figure*}[ht]
    \centering
    \includegraphics[width=1\textwidth]{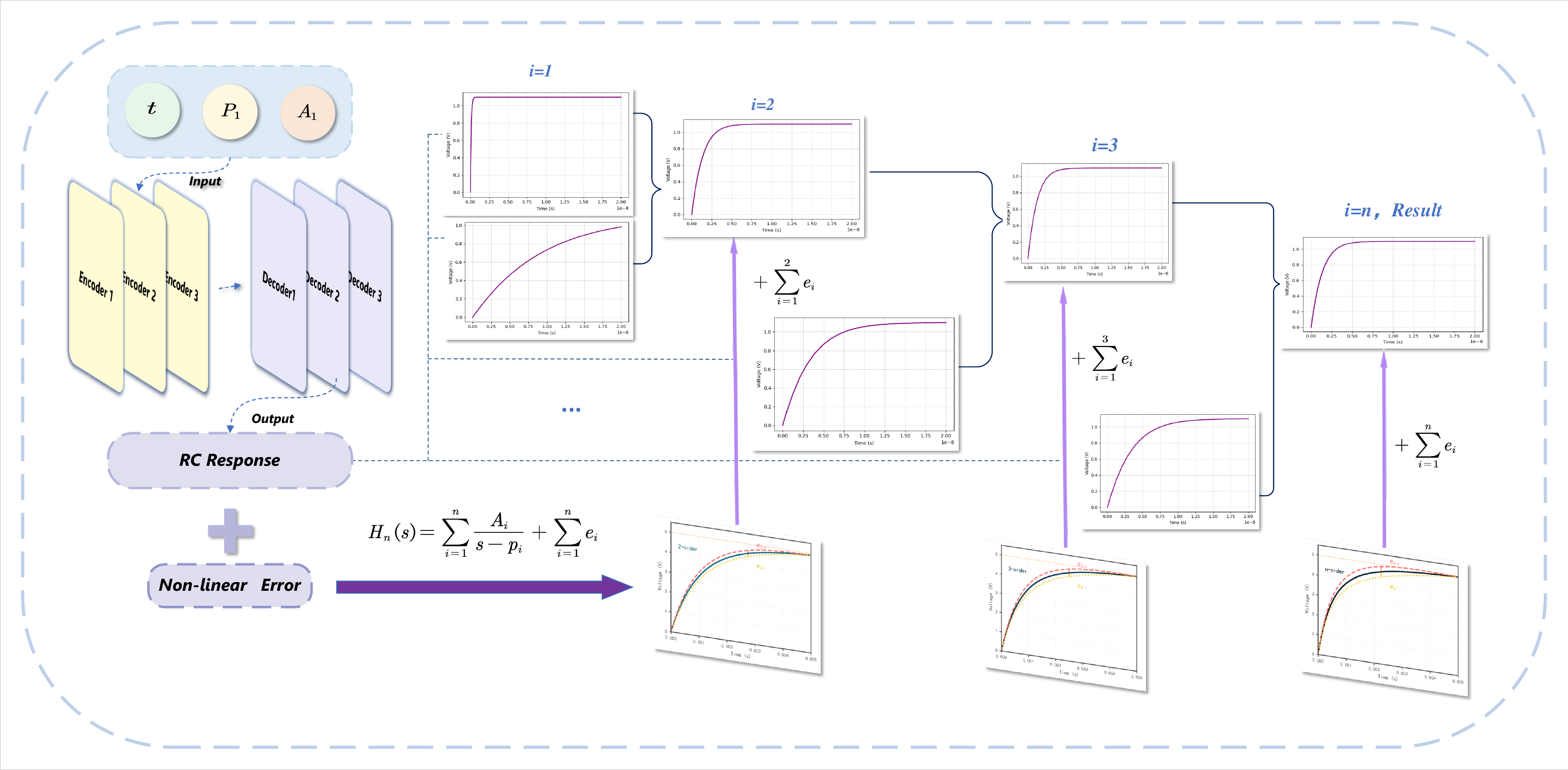}
    \caption{\textbf{Model Overview.} The model first constructs a baseline prediction by summing the first-order responses of pole-residue pairs. For each mode \( i \), the residual module \( e_{i} \) is trained using the current and previous poles, residues, and time information to correct the accumulated error. Residuals are added iteratively to refine the final prediction.
}
    \label{fig:model-overview}
\end{figure*}

The proposed model adopts a two-stage architecture: a base module first predicts individual first-order responses from partial fraction decomposition and aggregates them into a baseline output. A subsequent compensation network then iteratively refines this output by learning residual corrections across model orders. This overall architecture is illustrated in Figure~\ref{fig:model-overview}.

\section{Theoretical Background}

\subsection{\textbf{\textit{Decomposition Theorem for High-Order RC Network Transfer Functions}}}

Modeling nonlinear dynamical systems remains fundamental across many physical domains, where complex interactions often overlay intrinsic linear behaviors. In integrated circuit design, a prominent instance arises in interconnects, where nonlinear drivers interface with passive metal wires forming linear time-invariant (LTI) networks. The passive portion, shaped by parasitic resistances ($R_i$) and capacitances ($C_i$), governs signal delay, attenuation, and waveform distortion. The behavior of these RC networks is typically characterized in the Laplace domain via a transfer function $H(s)$ that maps input signals to output responses \cite{odabasioglu2003prima, feldmann2002efficient, antoulas2005approximation}. For a network with $m$ independent energy-storage elements, the transfer function takes the form shown in Equation (1):

\begin{equation}
H(s)=\frac{N(s)}{D(s)}=\frac{b_0 + b_1 s + \cdots + b_{m-1}s^{m-1}}{a_0 + a_1 s + \cdots + a_m s^m}.
\end{equation}
where the roots of the denominator $D(s)$ are the system poles $p_i = -1/\tau_i = -1/(R_i C_i)$, all located on the negative real axis due to the passive nature of RC circuits.

By the fundamental theorem of algebra and the Heine-Borel theorem, any strictly proper rational function with distinct poles admits a unique partial fraction expansion\cite{ogata2009modern} shown in Equation (2):
\begin{equation}
H(s)=\sum_{i=1}^m \frac{r_i}{s - p_i},\quad r_i=\left.\frac{N(s)}{D'(s)}\right|_{s=p_i}.
\end{equation}
This decomposition can be rigorously derived through two classical approaches. First, the method of undetermined coefficients constructs a linear system whose solution is guaranteed by the nonsingularity of the associated Vandermonde matrix when all poles are distinct. Second, the Cauchy residue theorem establishes the residue-based representation by integrating 
$H(s)$  around a closed contour enclosing all poles; analytic continuation then ensures the uniqueness of this expansion.

Further physical constraints arise from the realizability conditions of RC networks. All poles must lie strictly in the negative real domain to ensure overdamped and stable dynamics. While residues are typically real-valued, they may be either positive or negative, reflecting modal interference effects in higher-order coupled systems. Despite potential non-monotonicity at the modal level, the overall response remains physically consistent and interpretable, capturing the multi-timescale nature of signal propagation inherent to real interconnect behavior.

\subsection{\textbf{\textit{Generalization to Repeated Poles}}}
When a pole \(p_i\) of the transfer function has multiplicity \(k_i>1\) (with \(\sum_i k_i = m\)), the partial-fraction expansion naturally extends to form in Equation (3):
\begin{equation}
H(s) = \sum_{i=1}^q \sum_{j=1}^{k_i} \frac{r_{ij}}{(s - p_i)^j},
\label{eq:partial_fraction}
\end{equation}
where each higher-order residue is given by Equation (4): 
\begin{equation}
r_{ij} = \frac{1}{(k_i - j)!}
\left.\frac{d^{\,k_i - j}}{ds^{\,k_i - j}}\left[(s - p_i)^{k_i}H(s)\right]\right|_{s = p_i}.
\label{eq:residue_formula}
\end{equation}
In practice, however, exact repeated poles are rare in on-chip RC networks due to manufacturing tolerances and layout variations. Even when poles are nearly coincident, the corresponding higher-order residues \(r_{ij}\) tend to be small, and their time domain contributions \(t^{\,j-1}e^{p_i t}\) decay rapidly for \(p_i<0\). Consequently, one can safely ignore repeated-pole terms in most modeling tasks and rely on single-pole expansions to achieve high-fidelity simulations \cite{sheehan2007realizable}.

This comprehensive decomposition provides a unified framework for both time-domain and frequency-domain analysis of arbitrary high-order RC interconnect networks.

\subsection{Computational Complexity Analysis}

Traditional SPICE-based transient simulation begins by formulating a system of nonlinear equations based on circuit devices and Kirchhoff’s laws. Solving this system—typically via the Newton–Raphson method—is computationally expensive. For an $n$-node RC network, each iteration involves Jacobian construction and LU decomposition, resulting in a complexity of $\mathcal{O}(n^3)$. Even with sparse solvers, fill-in effects lead to an effective cost between $\mathcal{O}(n^{2.5})$ and $\mathcal{O}(n^3)$ over $T$ time steps and $P$ ports \cite{davis2006direct, demmel1997applied}.

In contrast, our method operates in the S-domain and eliminates matrix inversion. Computing the admittance or impedance to a single output node requires $\mathcal{O}(n)$ operations. Extending this to all $n$ nodes yields a total complexity of $\mathcal{O}(n^2)$, while avoiding iterative linear system solves.

\section{Data Acquisition and Preprocessing}
\subsection{Simulation Environment Setup}

The HSPICE simulation platform employing a 40-nm CMOS PDK ensured process-compliant device parameters ($V_{\mathrm{th}}$, $\lambda$, $I_{\mathrm{leak}}$), where a single-stage CMOS driver with IEEE 1481-2009-compliant RC networks (parasitics extracted via Python) was modeled through state space representation and converted to an $S$-domain transfer function; subsequent partial fraction decomposition yielded first-order subsystems characterized by poles $p_i$ and residues $r_i$ for neural operator-based time domain prediction.
\vspace{-0.1\baselineskip}

\subsection{Stimulation Configuration}
The input voltage waveform was configured as an ideal step signal (0 to VDD transition). Transient simulations covered both the signal rise phases (0 to 20 ns) and steady state behavior,with a time step resolution of 10ps.

\vspace{-0.5\baselineskip}
\subsection{Feature Representation and Supervision}
Each sample is defined by a tuple of conditioning inputs and supervised outputs. The conditioning inputs comprise a device type label, a sequence of transient time points \(t_1,\dots,t_T\), and a set of frequency domain features obtained via transfer function decomposition, encoded as pole–residue pairs \(\{(p_i,r_i)\}_{i=1}^m\). Together, these inputs capture the structural, temporal, and modal characteristics of the circuit.

The supervised target is the voltage response sequence $\{V_{\mathrm{out}}(t_1), \dots, V_{\mathrm{out}}(t_T)\}$, obtained from HSPICE simulation, which guides training via time-aligned regression.

\subsection{Feature Normalization}
To harmonize heterogeneous input features and enhance model robustness, we apply the following transformations in a single step shown in Equation (5):
\begin{align}
V'(t) &= \frac{V(t) - \min(V)}{\max(V) - \min(V)}, & 
t' &= \frac{\log_{10}(t) - \mu_t}{\sigma_t}
\end{align}
\vspace{-0.1\baselineskip}
where \(V(t)\) is the original voltage at time \(t\), \(\min(V)\) and \(\max(V)\) are its minimum and maximum over the waveform, \(t\) is a sampled transient time point, and \(\mu_t,\sigma_t\) are the mean and standard deviation of \(\log_{10}(t)\) across all samples. These normalizations place both features on comparable scales, mitigate the influence of outliers, and promote stable, efficient training.

\subsection{Transfer Function-Based RC Network Modeling}

We propose a compact, system-level modeling framework for standard-cell-driven RC interconnects by decomposing the Laplace-domain transfer function shown in Equation (6):

\begin{equation}
H(s)=\frac{V_{\mathrm{out}}(s)}{V_{\mathrm{in}}(s)}==\sum_{i=1}^{n}\frac{A_{i}}{\frac{s}{p_{i}}-1}
\end{equation}
where each decay rate \(p_{i}>0\) (inverse time constant) and residue \(A_{i}\in\mathbb{R}\) satisfies \(\sum_i A_i=1\), ensuring a normalized unit-step response.  This form captures the dominant exponential kernels \(e^{-p_i t}\) without explicit node-level modeling.

To enable neural‐network learning and generalization across circuits of varying size, we encode each mode as a pole-residue pair \((p_i, A_i)\), normalize all \(p_i\) and \(A_i\) by \(\max_i p_i\) and \(\max_i|A_i|\), sort pairs by descending \(|A_i|\). This interpretable mode sequence accurately reconstructs high-order RC responses with linear complexity and full spectral fidelity.

\section{Model Architecture Design}
\subsection{Baseline Module: First-Order Prediction}

The baseline module predicts the nonlinear voltage response of a single-mode RC system, specified by a pole-residue pair \((p, r)\), at discrete time points \(\{t_k\}_{k=1}^T\). Unlike ideal linear RC networks, the input waveform here first passes through nonlinear active components (e.g., CMOS drivers), making the overall system response analytically intractable. To address this, we employ a neural function approximator:$\hat{V}(t_k) = f_\theta(p, r, t_k)$.where \(f_\theta\) is a lightweight Transformer trained to capture the nonlinear mapping from modal and temporal inputs to voltage outputs\cite{vaswani2017attention}.

The model consists of three encoder and three decoder layers, each composed of multi-head self-attention, feed-forward sublayers with GELU activation, and layer normalization\cite{hendrycks2016gaussian}. Positional encoding is included to preserve temporal structure. Input features \((p, r, t_k)\) are embedded and processed in parallel to predict \(\hat{V}(t_k)\) at each time step. The network is trained end-to-end using the AdamW optimizer with weight decay, minimizing the mean squared error\cite{loshchilov2017decoupled}\cite{bengio2017deep} shown in Equation (7):

\begingroup
\setlength{\abovedisplayskip}{0.2em}
\setlength{\belowdisplayskip}{0.3em}
\setlength{\abovedisplayshortskip}{0.2em}
\setlength{\belowdisplayshortskip}{0.3em}
\begin{equation}
  \mathcal{L}_{\mathrm{MSE}} = \frac{1}{T} \sum_{k=1}^T 
  \bigl(\hat{V}(t_k) - V(t_k)\bigr)^2
\end{equation}
\endgroup
This architecture provides an accurate and generalizable first-order predictor that forms the foundation for modeling higher-order RC systems through residual correction.
\vspace{-0.5em}
\subsection{Compensation Module: Residual Correction}

The compensation module iteratively refines the baseline prediction by learning the residual error between successive model orders. For each order \(n\) and time point \(t_k\), the input feature vector is    $\bigl(n,\;p_n,\;r_n,\;p_{n-1},\;r_{n-1},\;t_k\bigr)$.where \((p_n,r_n)\) and \((p_{n-1},r_{n-1})\) denote the pole-residue pairs for the \(n\)th and \((n-1)\)th modes, respectively.The network outputs shown in Equation (8):

\begingroup
\setlength{\abovedisplayskip}{0.2em}
\setlength{\belowdisplayskip}{0.3em}
\setlength{\abovedisplayshortskip}{0.2em}
\setlength{\belowdisplayshortskip}{0.3em}
\begin{equation}
    \hat e_n(t_k)
    = g_\phi\bigl(n, p_n, r_n, p_{n-1}, r_{n-1}, t_k\bigr)
\end{equation}
\endgroup
which represents the corrective residual to be added to the order-\(n\) prediction at time \(t_k\).

The function \(g_\phi\) is implemented as a lightweight Transformer with the same depth and hyperparameters as the baseline module. It is trained end-to-end using the AdamW optimizer to minimize the residual mean squared error, thereby progressively correcting and refining higher-order predictions.

\subsection{Recursive Training Procedure}

Let \(\{(p_i, r_i)\}_{i=1}^n\) be the pole-residue pairs for the \(n\)th-order model and \(\{t_k\}_{k=1}^T\) the sampled time points. Denote by \(f_\theta\) the trained first-order predictor and by \(\{e_{\phi_j}\}_{j=1}^{n-1}\) the sequence of learned residual modules up to order \(n-1\). To train the \(n\)th residual module \(e_{\phi_n}\), we first accumulate the baseline prediction by summing \(f_\theta(p_i, r_i, t_k)\) for \(i = 1, \dots, n\). We then add all previously learned corrections \(e_{\phi_j}\) to form the current prediction and compute its discrepancy from the true output \(V_{\mathrm{out}}(t_k)\). This residual error serves as the target for \(e_{\phi_n}\), which is fit by minimizing the mean squared error over \(k = 1, \dots, T\) using the AdamW optimizer. Repeating this process for each order yields a cascade of lightweight modules that progressively refine the high-order RC response.

\begin{algorithm}[H]
\caption{Iterative Residual Correction Training}
\label{alg:residual-correction}
\begin{algorithmic}[1]

\Require Base predictor $f_\theta$, residual module set $\{e_{\phi_j}\}_{j=1}^N$, sampled data $\{t_k, V_{\mathrm{out}}(t_k)\}_{k=1}^T$
\Ensure Trained residual modules $\{e_{\phi_j}\}_{j=1}^N$

\State Initialize cumulative baseline prediction: $\hat{V}_{\mathrm{base}}(t_k) \gets 0,\ \forall k \in [1,T]$

\For{residual index $j = 1$ \textbf{to} $N$}
    \State Load current pole $p_j$ and residue $r_j$
    \Statex \textit{Phase 1: Baseline prediction update}
   
    \For{time step $k = 1$ \textbf{to} $T$}
        \State Update baseline: $\hat{V}_{\mathrm{base}}(t_k) \gets \hat{V}_{\mathrm{base}}(t_k) + f_\theta(p_j, r_j, t_k)$
    \EndFor
    \Statex \textit{Phase 2: Residual target computation}
    
    \For{time step $k = 1$ \textbf{to} $T$}
        \State Compute current prediction: 
        \[
        \hat{V}_j(t_k) \gets \hat{V}_{\mathrm{base}}(t_k) + \sum_{i=1}^{j-1} e_{\phi_i}(i, p_i, r_i, p_{i-1}, r_{i-1}, t_k)
        \]
        \State Compute residual target: 
        \[
        r_j(t_k) \gets V_{\mathrm{out}}(t_k) - \hat{V}_j(t_k)
        \]
    \EndFor
\Statex \textit{Phase 3: Module training}
    \State Build training set: $\mathcal{D}_j = \{(t_k, r_j(t_k))\}_{k=1}^T$
    \State Minimize the loss:
    \[
    \min_{\phi_j} \frac{1}{T} \sum_{k=1}^T \left( e_{\phi_j}(j, p_j, r_j, p_{j-1}, r_{j-1}, t_k) - r_j(t_k) \right)^2
    \]
    \State Update $\phi_j$ using gradient descent 
\EndFor

\end{algorithmic}
\end{algorithm}

By repeating this procedure for \(n=1,2,\dots,N\), we ensure each \(e_{\phi_n}\) learns to generalize the correction from order \(n\!-\!1\) to \(n\), yielding a cascade of residual models that together approximate the full high‑order response with minimal overfitting.Moreover, this recursive training scheme offers a degree of generalization. Each residual module \(e_{\phi_n}\) depends only on the pole–residue pairs of two adjacent orders and the current time point, without requiring knowledge of the full network topology or total number of nodes. As a result, the trained modules can be reasonably extended to refine predictions for moderately higher-order systems beyond those seen during training.

\subsection{Inference Procedure}

In the inference stage, we compute a single forward-pass estimate of the output waveform by first assembling a baseline response and then applying all residual corrections in parallel. Specifically, for each time sample \(t_k\) shown in Equation (9):
\vspace{-0.5\baselineskip}
\begin{equation}
\hat{V}_{\mathrm{base}}(t_k) = \sum_{i=1}^N f_\theta(p_i, r_i, t_k)
\end{equation}
\vspace{-0.5\baselineskip}

where \(p_i\) is the \(i\)th pole (inverse time constant) of the transfer function, \(r_i\) is the corresponding residue (modal weight), and \(f_\theta(p_i, r_i, t_k)\) denotes the base predictor’s output, typically \(r_i e^{-p_i t_k}\).

All \(N\) residual modules \(e_{\phi_j}\) are then evaluated and summed in parallel shown in Equation (10):
\vspace{-0.5\baselineskip}
\begin{equation}
\hat{V}_N(t_k) = \hat{V}_{\mathrm{base}}(t_k) + \sum_{j=1}^N e_{\phi_j}\bigl(j, p_j, r_j, p_{j-1}, r_{j-1}, t_k \bigr)
\end{equation}
where \(e_{\phi_j}(\cdot)\) is the \(j\)th trained residual correction module, and its inputs \((j, p_j, r_j, p_{j-1}, r_{j-1}, t_k)\) include the current and previous pole-residue pairs as well as the time \(t_k\).

This procedure yields the final prediction \(\hat{V}_N(t_k)\) at each \(t_k\) with \(\mathcal{O}(N)\) complexity, running 5--10\(\times\) faster than commercial tools like HSPICE by requiring just a single pass through the base predictor and residual modules.

\section{Experiment}

\subsection{Single‐Pole Transfer Function: Training and Test Performance}
We first evaluate the performance of our model on single‐pole transfer functions. The model achieves near‐perfect fit on the training data and strong generalization to unseen single‐pole functions.

\begin{figure}[htbp]
  \centering
  \captionsetup[subfigure]{justification=centering}

  \begin{subfigure}[t]{0.48\textwidth}  
    \includegraphics[width=\linewidth]{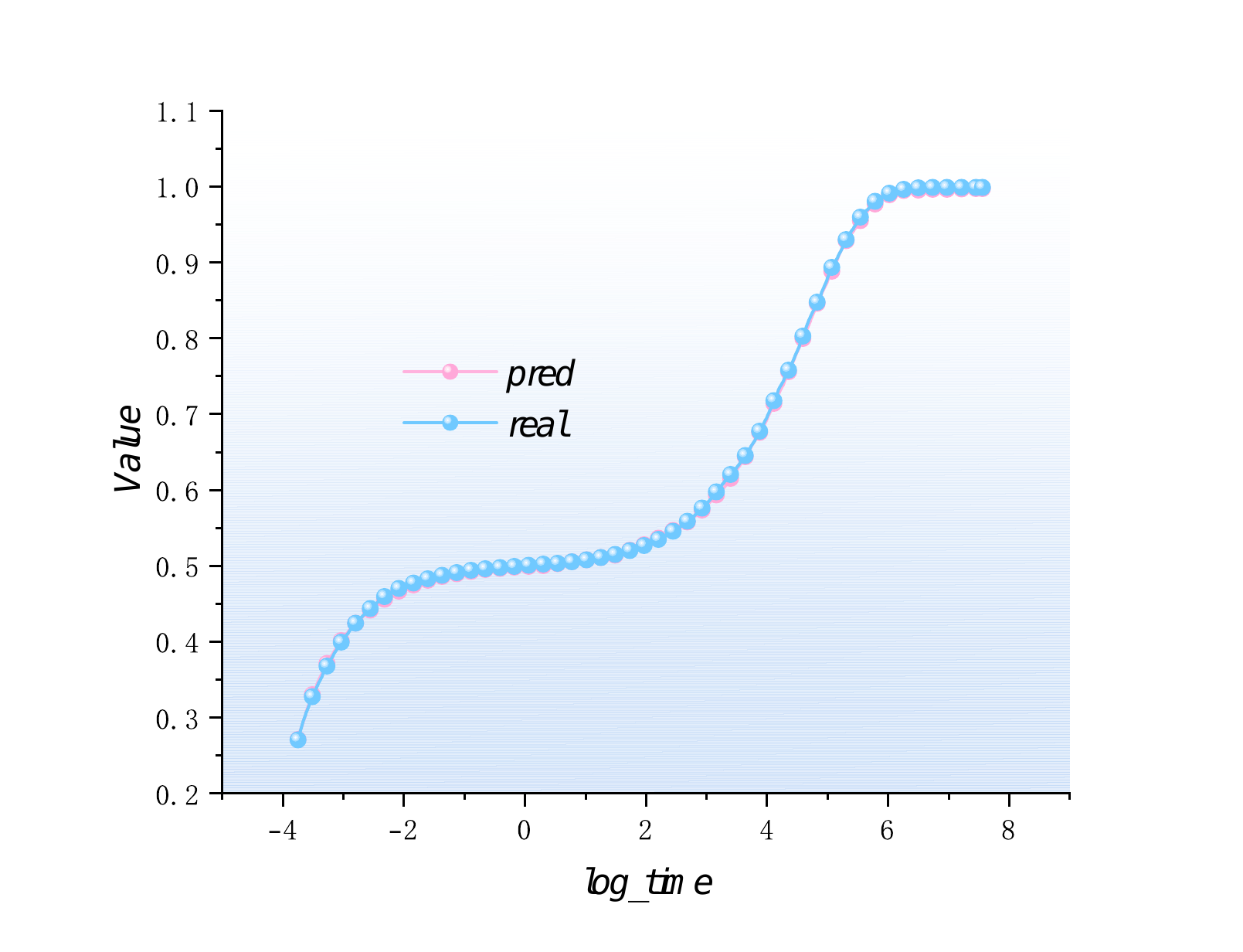}
    \caption{Single-Pole Transfer Function: Training vs.\ test performance. \(R^2=0.999\) illustrates near-perfect fit on the training data and strong generalization on held-out single-pole examples.}
    \label{fig:single_pole_performance}
  \end{subfigure}
  \hfill
  \begin{subfigure}[t]{0.48\textwidth}  
    \includegraphics[width=\linewidth]{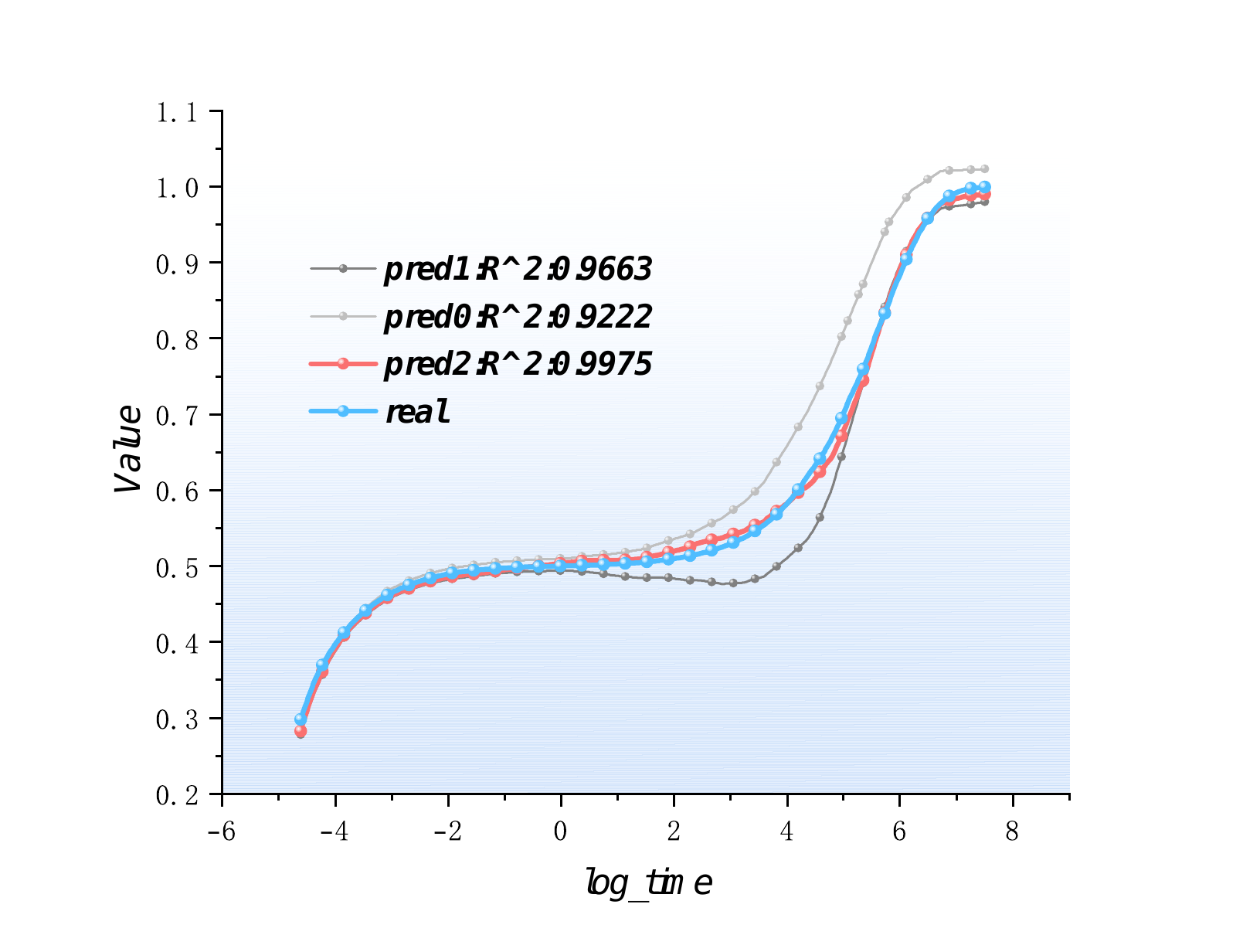}
    \caption{Error-Correction Model: This figure illustrates the step-by-step waveform correction process, from the raw prediction without any error correction through the successive application of first, second, and third-order error modules, culminating in a final fit of $R^2 = 0.9975$ }
    \label{fig:error_corr_comparison}
  \end{subfigure}

  \caption{(a) Model performance on single-pole transfer functions, showing minimal overfitting and excellent test-set accuracy. (b) Effectiveness of our recursive error-correction module on a three-pole example.}
  \label{fig:comparison_two_models}
\end{figure}

\subsection{Effectiveness of the Error-Correction Model}
To demonstrate the benefit of our recursive error-correction module, we evaluate on a three-pole transfer function example shown in Equation (11):
\vspace{-0.5\baselineskip}
\begin{equation}
  H(s) \;=\; \sum_{i=1}^{3} \frac{A_i}{\frac{s}{p_i} -1 }
\end{equation}

\vspace{-0.1\baselineskip}
where $\{p_i\}$ and $\{A_i\}$ are chosen such that the poles are well separated. Figure~\ref{fig:comparison_two_models} shows the predicted and true step responses over time. We compute the coefficient of determination
and obtain $R^2 = 0.9975$, confirming that the error-correction stage significantly improves accuracy over the base model.


\subsection{Generalization to Higher-Order Transfer Functions}
Next, we test the model’s ability to generalize to orders beyond those seen during training. We train exclusively on datasets of order up to 3 and then evaluate on transfer functions of orders 4 through 9. Table~\ref{tab:high_order_generalization} reports MSE and $R^2$ on each higher‐order test set. Despite never having seen orders above 3, the model retains strong predictive power, demonstrating effective extrapolation.

\begin{table}[htbp]
  \centering
  \captionsetup{skip=1em}  
  \begin{tabular}{ccccccc}
    \toprule
    Order & 4 & 5 & 6 & 7 & 8 & 9 \\
    \midrule
    $R^2$ & 0.986 & 0.997 & 0.995 & 0.990 & 0.979 & 0.964 \\
    \bottomrule
  \end{tabular}
  \caption{Generalization performance on higher-order transfer functions (trained on orders $\le3$).}
  \label{tab:high_order_generalization}
\end{table}

\begin{figure}[htbp]
  \centering
  \captionsetup[subfigure]{justification=centering}
  \begin{subfigure}[b]{0.3\textwidth}
    \includegraphics[width=\linewidth]{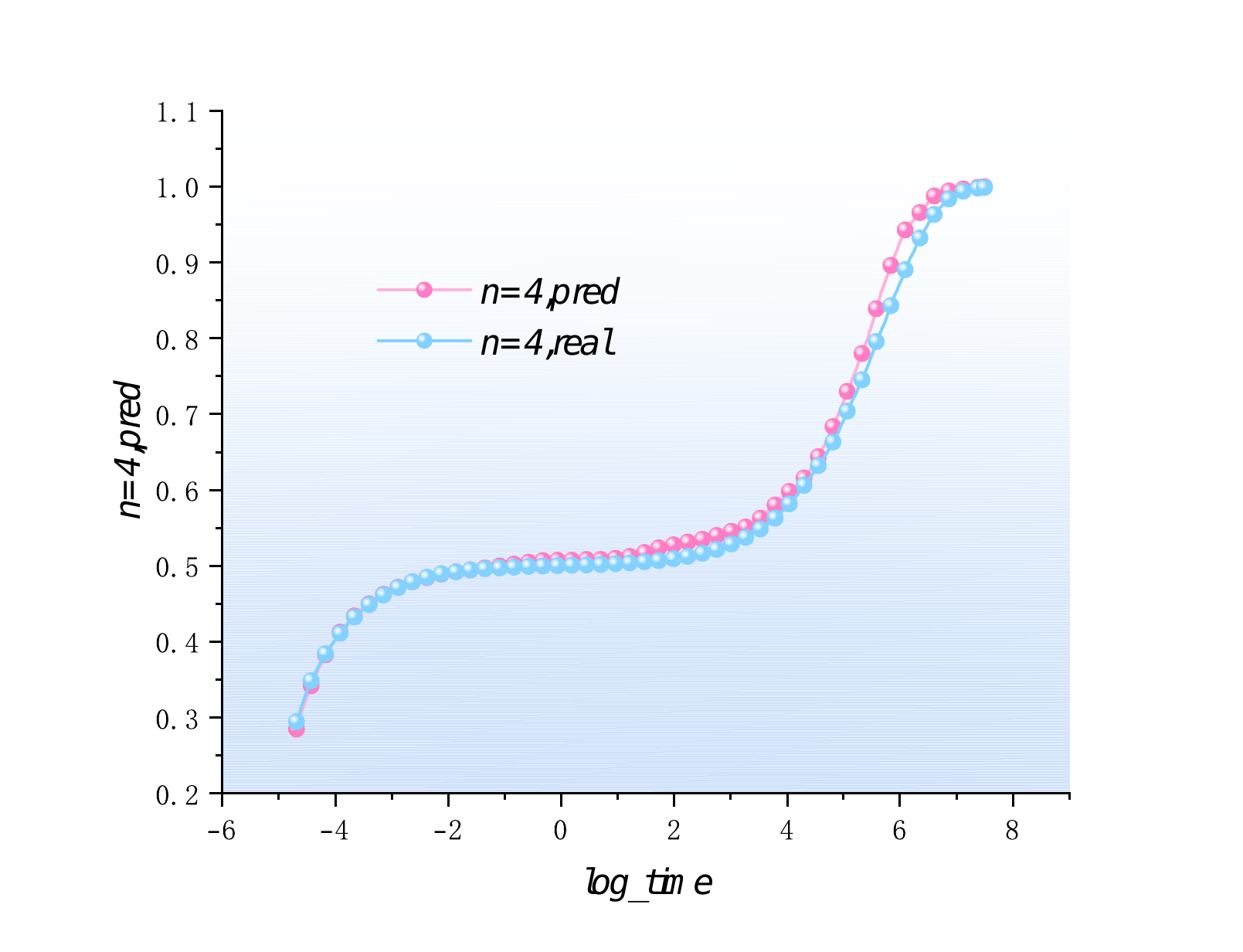}
    \caption{Order 4}
    \label{fig:pred4}
  \end{subfigure}
  \hfill
  \begin{subfigure}[b]{0.3\textwidth}
    \includegraphics[width=\linewidth]{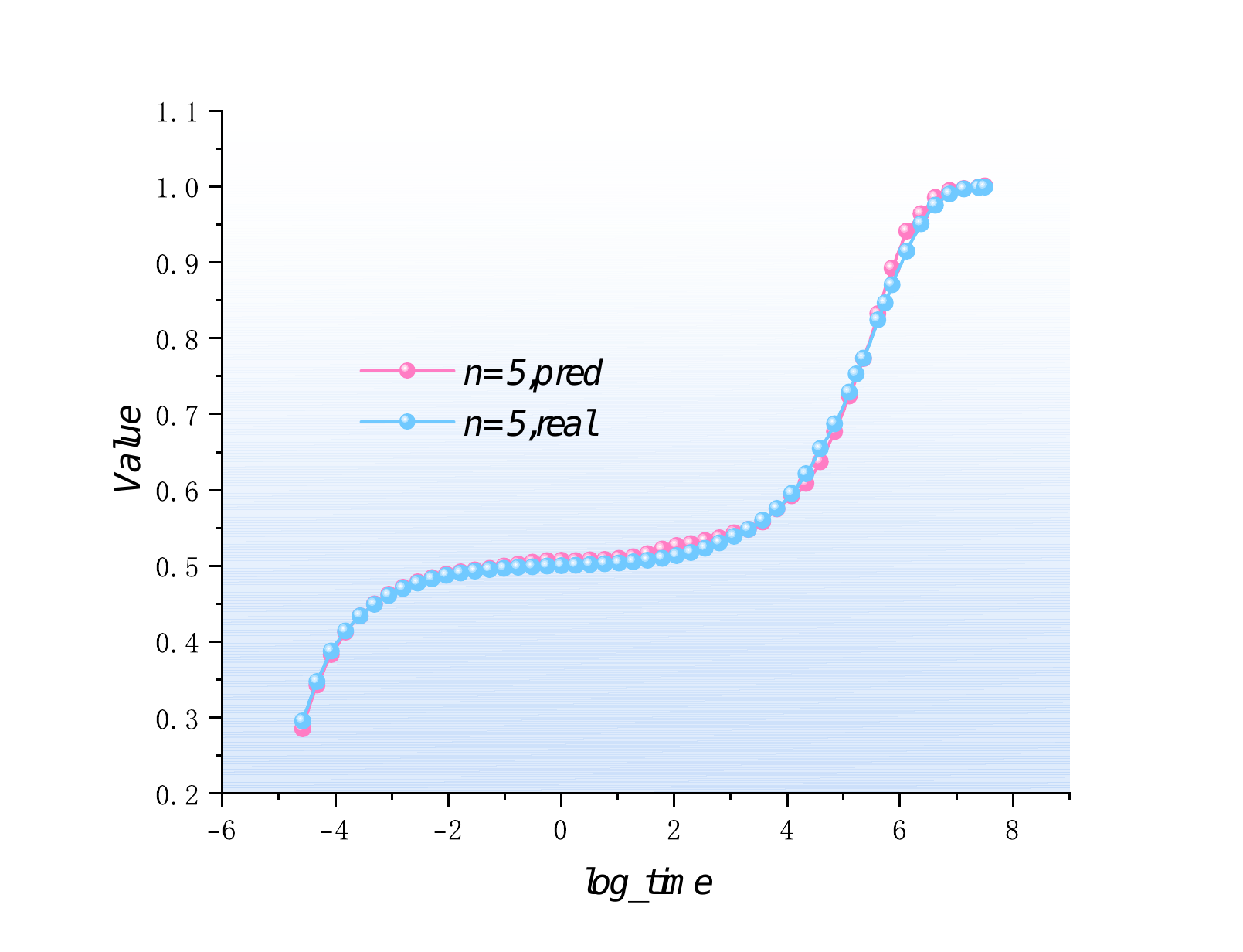}
    \caption{Order 5}
    \label{fig:pred5}
  \end{subfigure}
  \hfill
  \begin{subfigure}[b]{0.3\textwidth}
    \includegraphics[width=\linewidth]{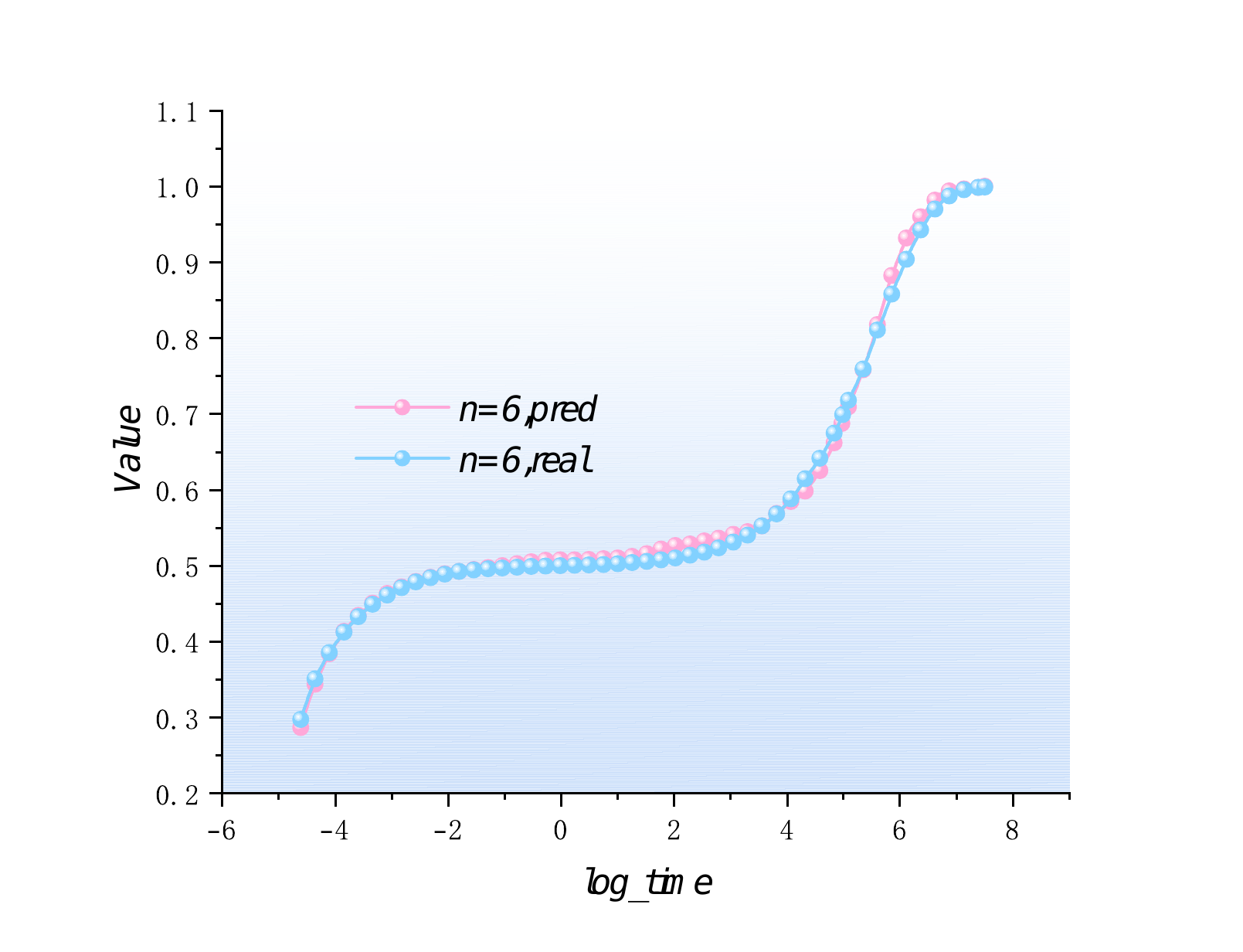}
    \caption{Order 6}
    \label{fig:pred6}
  \end{subfigure}

  \vspace{1em}

  \begin{subfigure}[b]{0.3\textwidth}
    \includegraphics[width=\linewidth]{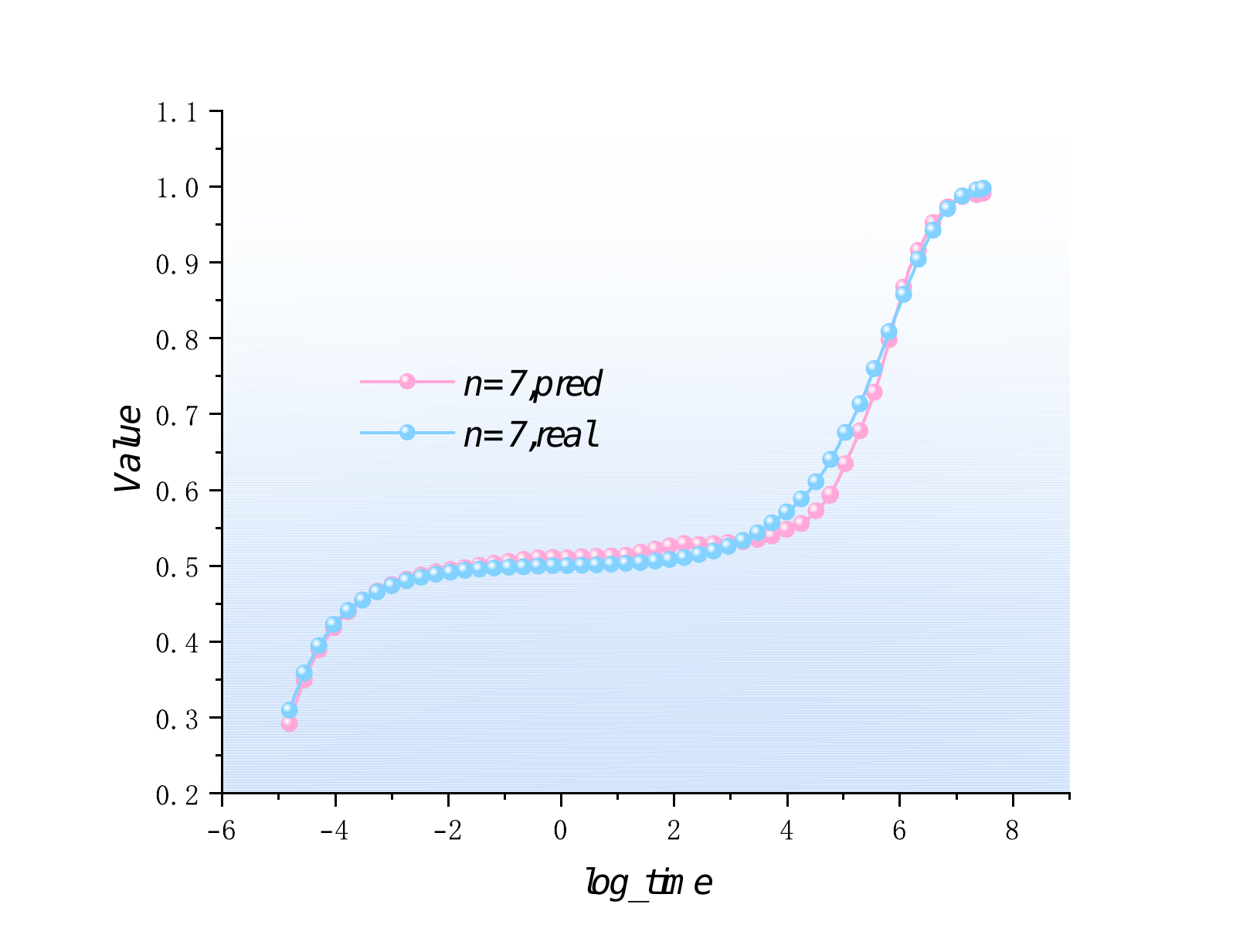}
    \caption{Order 7}
    \label{fig:pred7}
  \end{subfigure}
  \hfill
  \begin{subfigure}[b]{0.3\textwidth}
    \includegraphics[width=\linewidth]{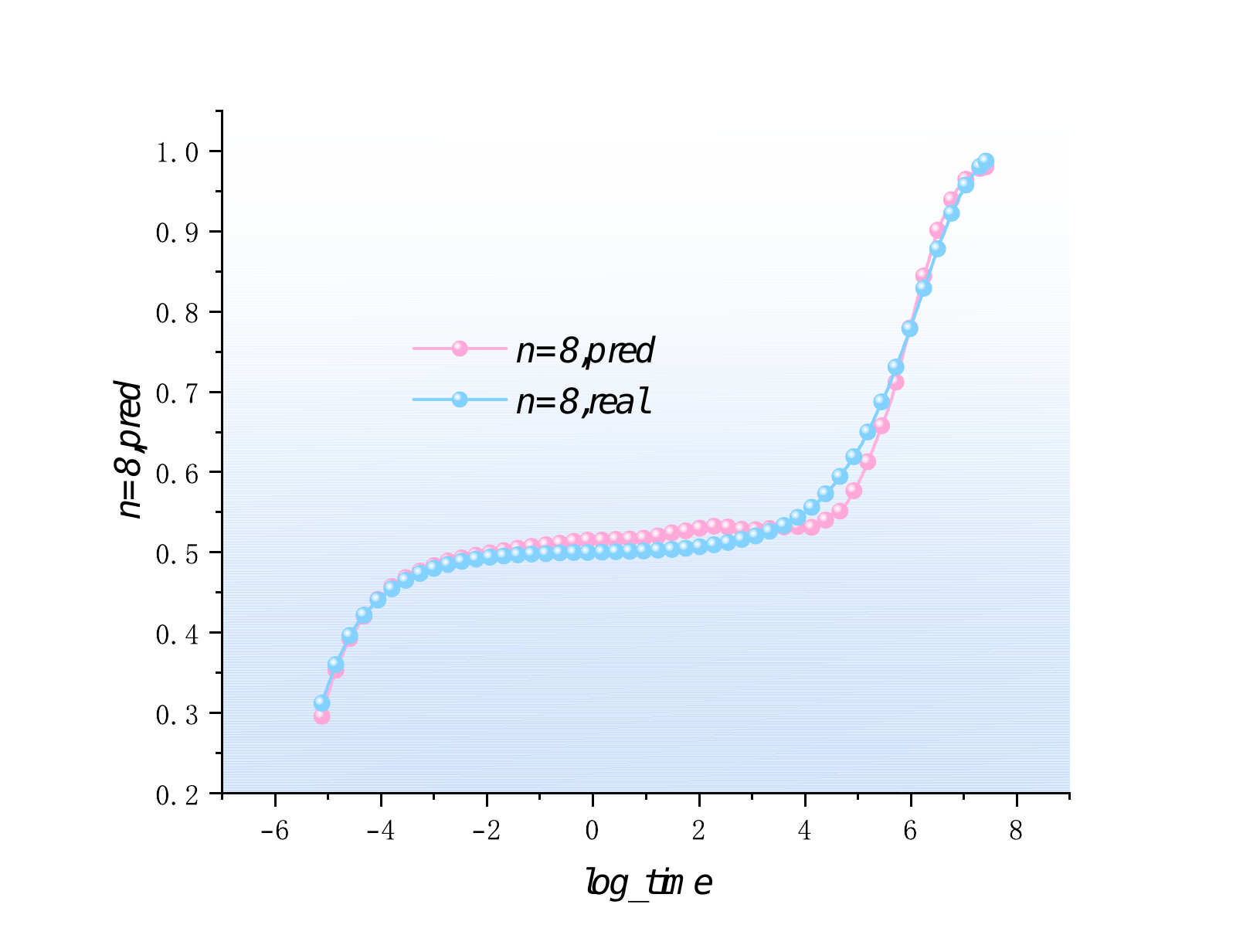}
    \caption{Order 8}
    \label{fig:pred8}
  \end{subfigure}
  \hfill
  \begin{subfigure}[b]{0.3\textwidth}
    \includegraphics[width=\linewidth]{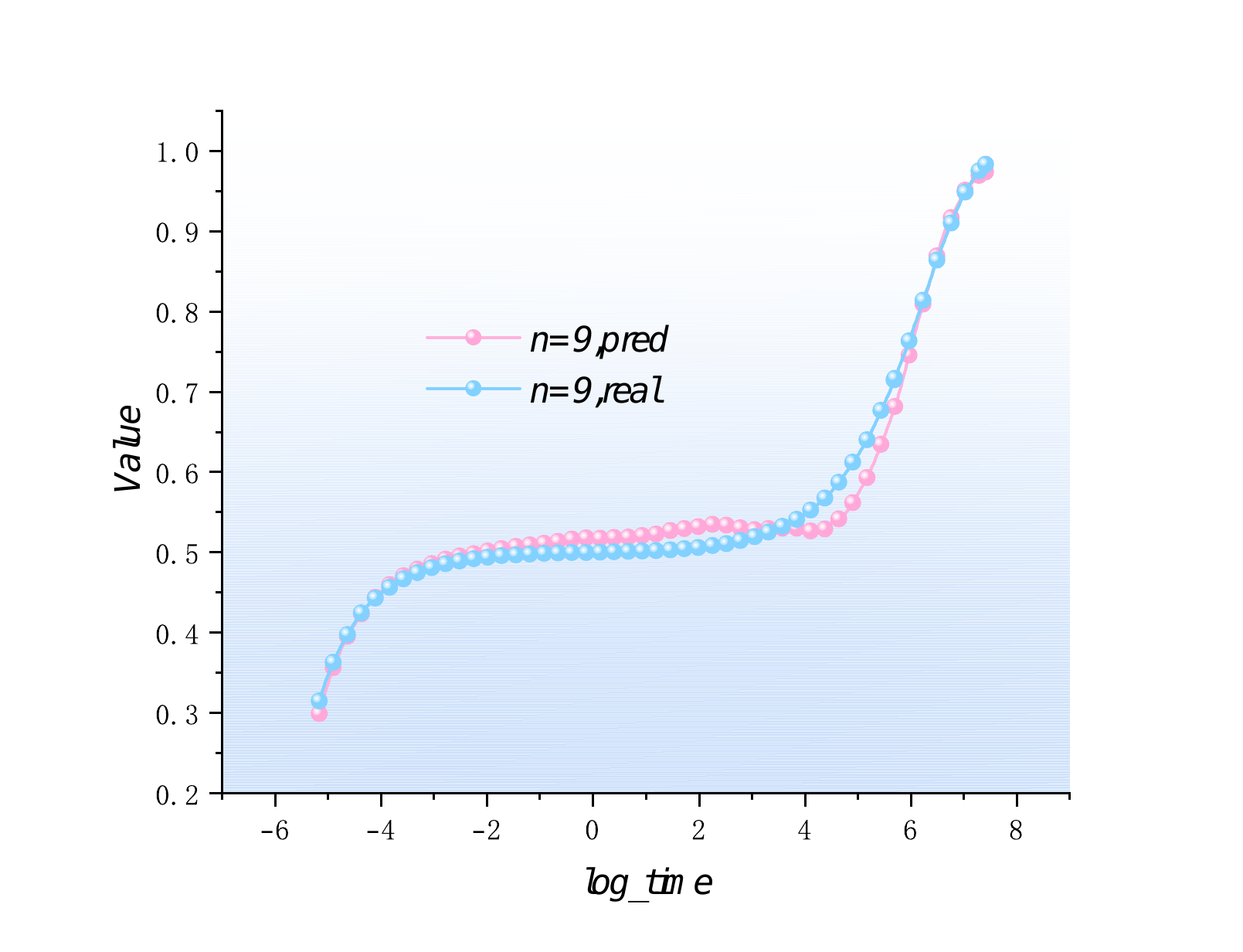}
    \caption{Order 9}
    \label{fig:pred9}
  \end{subfigure}

  \caption{Prediction vs.\ true response for transfer‐function orders 4 to 9.}
  \label{fig:prediction_comparisons}
\end{figure}

\subsection{Inference Time Comparison}
Finally, we compare the runtime of our S-Crescendo model against HSPICE on a 10ns transient simulation sampled at 1000 time steps. HSPICE runtimes for orders 1 through 10 were measured on a CPU node equipped with an AMD EPYC™ 7763 64-core processor and 256 GB of DDR4 RAM; S-Crescendo inference times were recorded on a workstation with an NVIDIA RTX 4090 GPU. Table~\ref{tab:runtime_comparison} reports the average simulation and inference times. Across all orders, S-Crescendo achieves more than two orders of magnitude speedup while preserving high accuracy.

\begin{table}[ht]
  \centering
  \caption{Runtime comparison between HSPICE and S-Crescendo across transfer function orders (10ns, 1000 steps).}
  \label{tab:runtime_comparison}
  \resizebox{\textwidth}{!}{%
  \begin{tabular}{l*{10}{c}}
    \toprule
    Order               & 1     & 2     & 3     & 4     & 5     & 6     & 7     & 8     & 9     & 10    \\
    \midrule
    HSPICE (s)          & 0.26  & 0.21  & 0.23  & 0.22  & 0.27  & 0.26  & 0.28  & 0.23  & 0.23  & 0.26  \\
    S-Crescendo (s)     & 0.014 & 0.019 & 0.022 & 0.018 & 0.023 & 0.028 & 0.031 & 0.034 & 0.039 & 0.042 \\
    Speedup (X)         & 18.6  & 11.1  & 10.5  & 12.2  & 11.7  & 9.3   & 9.0   & 6.8   & 5.9   & 6.2   \\
    \bottomrule
  \end{tabular}%
  }
\end{table}

\section{Further Discussion}

\subsection{Degradation of Accuracy at Higher Orders}  
While S-Crescendo performs well on low to mid-order transfer functions, its $R^2$ degrades as order $m$ increases. This is due to recursive error accumulation: each correction module $\varepsilon_k(t)$ adjusts not only current residuals but also propagates previous errors. If the single-pole state space has size $N$, then the full $m$-pole space grows as $N^m$, whereas each error model sees only $N$ samples during training. Thus, the fraction of covered states is shown in Equation (12):
\begin{equation}
    \frac{N \times m}{N^m}
\end{equation}
which shrinks rapidly with $m$, leading to sparse supervision and compounded inaccuracies.

\subsection{Reducing Data Dependency via Blocked Recursion}  
To alleviate error accumulation and data explosion, we propose a blocked recursion strategy. Instead of training a separate module per order, we group adjacent orders into blocks—for example, a shared module for orders 2–4, another for 5–6, etc. Each block is supervised on $O(N \times B)$ states (for block size $B$), yet extrapolates across $B$ orders. This reduces the number of recursive calls and lowers training demands, improving both runtime and scalability.

\subsection{Scaling of Inference Latency with Order}  
S-Crescendo's inference latency scales roughly linearly with order $m$, since each pole adds a forward pass. In contrast, tools like HSPICE collapse high-order dynamics via model reduction, maintaining near-constant runtime. To bridge this gap, we consider (i) collapsing low-impact poles into aggregate corrections, or (ii) integrating model-order reduction into the learned pipeline—both strategies aim to extend our efficiency gains to large-scale systems.

\subsection{Modeling Limitation: Repeated Poles in Transfer Functions}  
The model currently does not handle repeated poles, which introduce higher-order temporal terms like $t^k e^{\lambda t}$. To address this, future models can extend input features to include pole multiplicity, enabling learning from triplets $(p_i, A_i, m_i)$. This would broaden the model's applicability to more complex, higher-order dynamics.

\subsection{Outlook: Toward General Nonlinear–Linear Hybrid Systems}

Beyond RC modeling, the proposed framework extends naturally to hybrid systems with a ``nonlinear front-end + linear dynamic core'' architecture, common across engineering domains. Examples include switching converters in power electronics, where nonlinear control drives linear filters; analog front-ends, where transistor drivers interface with RC loads; and neural membrane models coupling nonlinear ion channels to capacitive elements. By modeling linear dynamics via Laplace-domain priors and learning nonlinear corrections, the framework enables efficient, interpretable emulation. Future directions include integrating operator learning or domain-specific constraints to broaden applicability.

\bibliographystyle{unsrtnat}
\bibliography{main.bbl}


\appendix

\section{Technical Appendices and Supplementary Material}
Technical appendices with additional results, figures, graphs and proofs may be submitted with the paper submission before the full submission deadline (see above), or as a separate PDF in the ZIP file below before the supplementary material deadline. There is no page limit for the technical appendices.


\newpage
\section*{NeurIPS Paper Checklist}


\begin{enumerate}

\item {\bf Claims}
    \item[] Question: Do the main claims made in the abstract and introduction accurately reflect the paper's contributions and scope?
    \item[] Answer: \answerYes{} 
    \item[] Justification: The abstract and introduction clearly state the key contributions of the paper.

\item {\bf Limitations}
    \item[] Question: Does the paper discuss the limitations of the work performed by the authors?
    \item[] Answer: \answerYes{} 
    \item[] Justification: The paper discusses limitations including error accumulation and explosive data requirements of this work in the beginning of Section 6, and provides further explanations and solutions in the rest of Section 6.

\item {\bf Theory assumptions and proofs}
    \item[] Question: For each theoretical result, does the paper provide the full set of assumptions and a complete (and correct) proof?
    \item[] Answer: \answerYes{} 
    \item[] Justification: We provide the full set of assumptions for each theoretical result, with complete and correct proofs included in the appendix.

    \item {\bf Experimental result reproducibility}
    \item[] Question: Does the paper fully disclose all the information needed to reproduce the main experimental results of the paper to the extent that it affects the main claims and/or conclusions of the paper (regardless of whether the code and data are provided or not)?
    \item[] Answer: \answerYes{}{} 
    \item[] Justification:  We will release both our training and inference code, along with detailed instructions for constructing the dataset used in our experiments. While the actual dataset cannot be released due to constraints, all necessary steps and scripts will be provided to enable reproduction of the main results.

\item {\bf Open access to data and code}
    \item[] Question: Does the paper provide open access to the data and code, with sufficient instructions to faithfully reproduce the main experimental results, as described in supplemental material?
    \item[] Answer: \answerYes{}{} 
    \item[] Justification:   We will release our training and inference code on GitHub. The model and dataset will not be open-sourced, but we provide the detailed procedure for constructing the dataset. We partially release the dataset generation scripts, and the data preprocessing steps are described in the paper.


\item {\bf Experimental setting/details}
    \item[] Question: Does the paper specify all the training and test details (e.g., data splits, hyperparameters, how they were chosen, type of optimizer, etc.) necessary to understand the results?
    \item[] Answer: \answerYes{} 
    \item[] Justification:We will include full training and test details in the supplementary material.These details are sufficient to understand and reproduce the experimental results.

\item {\bf Experiment statistical significance}
    \item[] Question: Does the paper report error bars suitably and correctly defined or other appropriate information about the statistical significance of the experiments?
    \item[] Answer: \answerYes{} 
    \item[] Justification: The statistical significance of our experiments is evaluated by comparing the model's predictions with the gold-standard results generated by HSPICE simulations. We report root mean square error ($R^2$) as the primary metric to quantify the accuracy of our predictions. This provides a reliable and consistent measure of model performance.

\item {\bf Experiments compute resources}
    \item[] Question: For each experiment, does the paper provide sufficient information on the computer resources (type of compute workers, memory, time of execution) needed to reproduce the experiments?
    \item[] Answer: \answerYes{} 
    \item[] Justification: All experiments were conducted on a single NVIDIA RTX 4090 GPU platform with 32 GB of memory. Training was performed on this platform, and the training time varied depending on the order of the system and the dataset size. Inference times are reported and compared in the main paper. The setup ensures that the experiments are reproducible on similar hardware.
    
\item {\bf Code of ethics}
    \item[] Question: Does the research conducted in the paper conform, in every respect, with the NeurIPS Code of Ethics \url{https://neurips.cc/public/EthicsGuidelines}?
    \item[] Answer: \answerYes{}

\item[] Justification: We have reviewed the NeurIPS Code of Ethics and confirm that our work fully complies with it. Our research does not involve human data, privacy concerns, or potential for misuse. All results are reported transparently and responsibly.


\item {\bf Broader impacts}
    \item[] Question: Does the paper discuss both potential positive societal impacts and negative societal impacts of the work performed?
    \item[] Answer: \answerNA{} 
    \item[] Justification: This work is a technical study focused on physical modeling and simulation of high-order nonlinear systems to improve efficiency and accuracy in VLSI design. It has no direct societal applications and therefore does not discuss societal impacts. The method is primarily for engineering simulations with negligible risk of misuse or social issues like fairness or privacy.

    
\item {\bf Safeguards}
    \item[] Question: Does the paper describe safeguards that have been put in place for responsible release of data or models that have a high risk for misuse (e.g., pretrained language models, image generators, or scraped datasets)?
    \item[] Answer: \answerNA{} 
    \item[] Justification:This work is a technical simulation study and does not involve high-risk models or datasets that require special safeguards.

\item {\bf Licenses for existing assets}
    \item[] Question: Are the creators or original owners of assets (e.g., code, data, models), used in the paper, properly credited and are the license and terms of use explicitly mentioned and properly respected?
    \item[] Answer: \answerNA{} 
    \item[] Justification:We utilized Synopsys HSPICE (U-2023.03-SP2-2) under a valid institutional license for circuit simulations. Additionally, we employed open-source Python libraries with proper license acknowledgments provided in the supplemental materials.

\item {\bf New assets}
    \item[] Question: Are new assets introduced in the paper well documented and is the documentation provided alongside the assets?
    \item[] Answer: \answerYes{}{} 
    \item[] Justification:  We will release the relevant code and the dataset generation process alongside the paper. Detailed documentation, including usage instructions and experimental setups, will be provided in the supplementary material.

    \begin{itemize}
        \item The answer NA means that the paper does not release new assets.
        \item Researchers should communicate the details of the dataset/code/model as part of their submissions via structured templates. This includes details about training, license, limitations, etc. 
        \item The paper should discuss whether and how consent was obtained from people whose asset is used.
        \item At submission time, remember to anonymize your assets (if applicable). You can either create an anonymized URL or include an anonymized zip file.
    \end{itemize}

\item {\bf Crowdsourcing and research with human subjects}
    \item[] Question: For crowdsourcing experiments and research with human subjects, does the paper include the full text of instructions given to participants and screenshots, if applicable, as well as details about compensation (if any)? 
    \item[] Answer: \answerNA{} 
    \item[] Justification: This work does not involve crowdsourcing or research with human subjects.

\item {\bf Institutional review board (IRB) approvals or equivalent for research with human subjects}
    \item[] Question: Does the paper describe potential risks incurred by study participants, whether such risks were disclosed to the subjects, and whether Institutional Review Board (IRB) approvals (or an equivalent approval/review based on the requirements of your country or institution) were obtained?
    \item[] Answer: \answerNA{} 
    \item[] Justification: This work does not involve crowdsourcing or research with human subjects.

\item {\bf Declaration of LLM usage}
    \item[] Question: Does the paper describe the usage of LLMs if it is an important, original, or non-standard component of the core methods in this research? Note that if the LLM is used only for writing, editing, or formatting purposes and does not impact the core methodology, scientific rigorousness, or originality of the research, declaration is not required.
    \item[] Answer: \answerNA{} 
    \item[] Justification: The research uses Transformer architecture but does not involve large language models (LLMs) as core components.
    \begin{itemize}
        \item The answer NA means that the core method development in this research does not involve LLMs as any important, original, or non-standard components.
        \item Please refer to our LLM policy (\url{https://neurips.cc/Conferences/2025/LLM}) for what should or should not be described.
    \end{itemize}

\end{enumerate}

\end{document}